\documentclass[pra,nobibnotes,aps,twocolumn,showpacs,superscriptaddress,amsfonts,amsmath,floatfix]{revtex4-1}

\usepackage[capitalize]{cleveref}
\usepackage{graphicx}
\usepackage{color}
\usepackage{import}
\usepackage{wrapfig}
\usepackage{mathtools}
\usepackage{bm}
\usepackage{amsmath}
\usepackage{amsfonts}
\usepackage{amssymb}

\begin{document}


\title{Modulation instability in the nonlinear Schr{\"o}dinger equation with a synthetic magnetic field: gauge matters}

\author{Karlo Lelas}
\email{klelas@ttf.unizg.hr}
\affiliation{Faculty of Textile Technology, University of Zagreb, Prilaz baruna Filipovi\'{c}a 28a, 10000 Zagreb, Croatia}

\author{Ozana \v{C}elan}
\affiliation{Department of Physics, Faculty of Science, University of Zagreb, Bijeni\v{c}ka c. 32, 10000 Zagreb, Croatia}
\author{David Prelogovi\'c}
\affiliation{Scuola Normale Superiore, Piazza dei Cavalieri 7, I-56126 Pisa, Italy}
\author{Hrvoje Buljan}
\affiliation{Department of Physics, Faculty of Science, University of Zagreb, Bijeni\v{c}ka c. 32, 10000 Zagreb, Croatia}

\author{Dario Juki\'c}
\email{djukic@grad.unizg.hr}
\affiliation{Faculty of Civil Engineering, University of Zagreb, A. Ka\v{c}i\'{c}a  Mio\v{s}i\'{c}a 26, 10000 Zagreb, Croatia}

\date{\today}

\begin{abstract}
We theoretically investigate the phenomenon of modulation instability for systems obeying nonlinear
Schr{\"o}dinger equation, which are under the influence of an external homogeneous synthetic magnetic field.
For an initial condition, the instability is detected numerically by comparing dynamics with and without 
a small initial perturbation; the perturbations are characterized in a standard fashion by wavevectors in momentum space. 
We demonstrate that the region of (in)stability in momentum space, as well as time-evolution in real space, 
for identical initial conditions, depend on the choice of the gauge (i.e., vector potential) used to describe 
the homogeneous synthetic magnetic field.
This superficially appears as if the gauge invariance is broken, but this is not true. 
When the system is evolved from an identical initial condition in two different gauges, 
it is equivalent to suddenly turning on the synthetic magnetic field at $t=0$. 
This gives rise, via Faraday's law, to an initial instantaneous kick of a synthetic electric field to the wavepacket, 
which can differ for gauges yielding an identical uniform magnetic field at $t>0$. 
\end{abstract}

\maketitle

\section{Introduction}

Modulation instability (MI) is a nonlinear phenomenon which has been long studied in various physical systems including fluid dynamics, nonlinear optics, and plasma physics \cite{Lighthill1965, Whitham1965, Bespalov1966, Benjamin1967a, Benjamin1967b, Ostrovsky1967, Taniuti1968, Soljacic2000, Kip2000, Agrawal2013} (for historical overview of early work, see Ref.~\cite{Zakharov2009}).
Following major experimental developments with ultracold atomic gases, MI has been investigated for Bose-Einstein condensates (BECs)~\cite{Konotop2002, Smerzi2002, Catalioti2003, Theocharis2003}.
MI occurs when small (long-wavelength) perturbations on the uniform background intensity become exponentially amplified. In this way instability, which develops from the interplay of nonlinearity and dispersion~\cite{Agrawal2013}, breaks the symmetry of the uniform state.
As a recent example, experiments with ultracold atoms have investigated the role of MI in the formation of matter-wave solitons~\cite{Nguyen2017}, in analogy with extensive studies of solitons in optics~\cite{Chen2012}.
Here we explore MI in a nonlinear system with a synthetic magnetic field, which connects the research of nonlinear 
and topological phenomena.

The implementation of synthetic gauge fields is of great interest in atomic systems~\cite{Dalibard2011, Goldman2014}, 
because it can enable exploring topological phases of matter~\cite{Cooper2019}.
Analogous ideas on photonic platforms have led to the emergence of topological 
photonics~\cite{Lu2014, Ozawa2019}.
There is extensive literature on synthetic gauge fields and topological phases in these systems, 
as some of the ideas arose a quarter of century ago~\cite{Dum1996};
a number of comprehensive reviews on these topics~\cite{Dalibard2011, Goldman2014, Cooper2019, Lu2014, Ozawa2019}, 
some of which are very recent~\cite{Cooper2019, Ozawa2019}, have been published.

Let us mention a few of the schemes used for creation of synthetic gauge fields for 
atoms~\cite{Madison2000, AboShaeer2001, Lin2009, Aidelsburger2011, Struck2012, Miyake2013, 
Aidelsburger2013, Jotzu2014} and photons~\cite{Rechtsman2013, Rechtsman2013b, Hafezi2013, Yang2019}.
The first scheme for ultracold atoms was implemented in rapidly rotating BECs by employing 
the analogy between the Coriolis and the Lorentz force~\cite{Madison2000, AboShaeer2001}. 
The first implementation using light-atom interaction employed the analogy between 
the Aharonov-Bohm phase for charged particles, and the Berry phase 
for ultracold atoms with spatially dependent Raman coupling between internal hyperfine states~\cite{Lin2009}.
Very successful schemes were implemented in optical lattices~\cite{Aidelsburger2011, Struck2012, Miyake2013, 
Aidelsburger2013, Jotzu2014}, where the tunneling matrix element between neighboring sites 
is engineered to acquire a synthetic Peierls phase.

An equivalent strategy to engineer coupling between optical cavities, 
or photonic lattice sites, has been proposed in photonic systems, 
e.g., see Refs.~\cite{Umucalilar2011, Hafezi2011, Fang2012, Longhi2013, Dubcek2015}. 
It was successfully implemented by using link resonators of different length~\cite{Hafezi2013}, 
to image topological edge states~\cite{Hafezi2013}. 
A scheme mimicking strained graphene was used in photonic lattices to obtain artificial magnetic fields~\cite{Rechtsman2013}.
Interestingly, photonic Floquet topological insulators were implemented using helical waveguides 
which yield synthetic electric fields~\cite{Rechtsman2013b}. 
A non-Abelian gauge field gas been synthesized recently in an optical setup~\cite{Yang2019}.

A majority of work on synthetic gauge fields and topological phases are in noninteracting systems 
(single-particle phenomena) for ultracold atoms~\cite{Dalibard2011, Goldman2014, Cooper2019}, 
and in linear photonic systems~\cite{Lu2014, Ozawa2019}. However, when interactions or nonlinearity 
are turned on, intriguing phenomena such as the Fractional Quantum Hall Effect can emerge~\cite{Tsui1982,Laughlin1983}.
The nonlinear photonic phenomena addressed in the topological context include an 
analysis of the Hofstadter butterfly in a nonlinear Harper lattice~\cite{Manela2010}, 
solitons~\cite{Lumer2013, Leykam2016, Solnyshkov2017, Marzuola2019, Smirnova2019}, 
nonlinear harmonic generation~\cite{Kruk2019, Wang2019}, topological 
lasers~\cite{StJean2017, Bahari2017, Bandres2018}, topological transitions~\cite{Hadad2016, Bleu2016},
nonlinear control~\cite{Dobrykh2018}, and nonlinear pumping \cite{Bisianov2019} of topological edge states.

In this paper, we theoretically explore how the implementation of the homogeneous synthetic magnetic field in systems modeled by the two-dimensional (2D) nonlinear Schr{\"o}dinger equation (NLSE) affects the MI phenomenon.
Dynamics of weakly interacting BECs (in the mean-field approximation) 
and propagation of light through nonlinear media are both described by the NLSE~\cite{Pitaevskii2003, Agrawal2013}; 
for BECs it is usually referred to as the Gross-Pitaevskii equation (GPE)~\cite{Pitaevskii2003}.
Therefore, our study is applicable to both ultracold atomic and photonic systems.
In two dimensions, the addition of an external uniform magnetic field into a Hamiltonian leads to harmonic terms (among others) 
in the NLSE, resembling the scalar harmonic trap potential.
For this reason, in Sec. \ref{sect:2} we first outline the study of MI in one-dimensional (1D) harmonic traps, 
following the work in Ref.~\cite{Theocharis2003}. 
The NLSE in 2D with a magnetic field is introduced in Sec. \ref{sect:3} for different gauges of vector potential.
In Sec. \ref{sect:4} we numerically explore MI in 2D NLSE with the synthetic magnetic field.
More specifically, we explore the time evolution of an initial Thomas-Fermi profile wavepacket 
(with and without perturbations), and compare its dynamics for symmetrical and Landau gauges.
Perturbations are characterized in momentum space. 
We demonstrate that the dynamics of wavepackets with identical initial conditions, 
and regions of (in)stability in momentum space, are dependent on the choice of the gauge.
This may seem as if gauge invariance is broken, however, this is not true. 
When the system is evolved from an identical initial condition in two different gauges
yielding the same uniform synthetic magnetic field, it is equivalent to suddenly turning on the field at $t=0$ 
(or $z=0$ in spatial photonics). 
At this instance, fields arising from gauges differ and our results can be explained with Faraday's law: as the 
homogeneous synthetic magnetic field is turned on, an instantaneous kick of a synthetic electric field, 
which differs in the two gauges, occurs and affects subsequent dynamics.
This {\em gauge matters} effect has already been noted in Ref.~\cite{Spielman2015} in a different context. 
Finally, in Sec. \ref{sect:5} we conclude and summarize our results.

\section{Modulation instability in 1D NLSE with harmonic potential}
\label{sect:2}

We start by studying a 1D system in a harmonic potential which satistfies the NLSE,
\begin{equation}
i \frac{\partial \psi}{\partial t} = \left( -\frac{\partial^2}{\partial x^2} + a x^2 + \eta \left| \psi \right|^2 \right) \psi. 
\label{nlse1d}
\end{equation}
Here, $a$ is the harmonic oscillator constant, and $\eta < 0$ characterizes the strength of the nonlinearity.
Due to the harmonic potential, this equation does not have a homogeneous ground state for which a standard MI analysis could be performed.
This problem has been thoroughly studied in Ref.~\cite{Theocharis2003} numerically.
Mechanism of quasi-integrability of Eq. (\ref{nlse1d}) (with positive nonlinearity) was recently considered in Ref.~\cite{Malomed2018}.   

For clarity, we will briefly review demonstration of MI in this system.
We assume that the initial state is the ground state of the stationary NLSE in the Thomas-Fermi (TF) approximation.
The stationary equation reads
\begin{equation}
\mu \psi = \left( -\frac{\partial^2}{\partial x^2} + a x^2 + \eta' \left| \psi \right|^2 \right) \psi,
\end{equation}
where $\mu$ is the chemical potential, and the nonlinearity is positive, $\eta' >0$.
In the TF approximation the kinetic energy term is neglected, and the resulting wave function is $\psi_{TF} = \sqrt{\frac{\mu - ax^2}{\eta}}$ for $|x|<\sqrt{\mu/a}$, and $\psi_{TF}=0$ elsewhere.
We choose the chemical potential $\mu=1$, harmonic oscillator constant $a=0.0025$ and strength of the nonlinearity $\eta'=1$  \cite{Theocharis2003}.
Since MI is expected to occur at negative values of nonlinearity, we quench the system so that the sign of nonlinearity is switched from positive $\eta'$ to negative $\eta = -\eta'$ at $t=0$, and investigate the time-evolution of Eq.~\eqref{nlse1d} with the TF initial condition.
In the context of ultracold atomic gases, this quench in the nonlinearity can be achieved experimentally by using Feshbach resonances \cite{Chin2010}.

\begin{figure}
\includegraphics[width=0.48\textwidth]{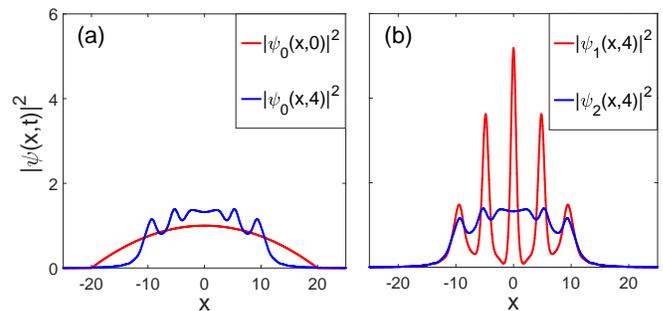}
\caption{Modulation instability in 1D NLSE with a harmonic potential. (a) Density of the initial state $\psi_0(x,0)=\psi_{TF}$ (red line) and density of the time-evolved state $\psi_0(x,4)$ (blue line). (b) Densities of the time-evolved states $\psi_1(x,4)$ (red line) and $\psi_2(x,4)$ (blue line), obtained with time-evolution of the perturbed initial states $\psi_1(x,0)=\mathcal{N}_1\psi_{TF}(1+0.01\cos x)$ and $\psi_2(x,0)=\mathcal{N}_2\psi_{TF}(1+0.01\cos 2x)$, respectively. See text for details.}
\label{fig:Fig1}
\end{figure}

The time-dynamics of the initial state $\psi_0(x,0)=\psi_{TF}$, which we observe after the quench, results in density modulations, shown in the Fig. \ref{fig:Fig1}(a) at $t=4$, but they are here present only due to the fact the TF state is not an eigenstate of the system.
%
%
On the other hand, MI is demonstrated by studying time-evolution of a slightly perturbed initial TF state. In our calculations, the perturbed initial states are of the form $\psi_k(x,0)=\mathcal{N}_{k}\psi_{TF}(1+\cos kx)$, where $k=\{1,2\}$, and the constant $\mathcal{N}_{k}$ ensures that both perturbed initial states have the same normalization as the unperturbed initial TF state. As visible in Fig. \ref{fig:Fig1}(b), adding an appropriate noise term with $k=1$ leads to density modulations which develop quickly in time, indicating that the initial state is unstable with respect to the perturbation with $k=1$. 
In contrast, the perturbation with $k=2$ does not destabilize the trajectory from the initial TF state, as visible in Fig. \ref{fig:Fig1}(b).

\section{Two-dimensional NLSE with a synthetic magnetic field: Which gauge to use?}
\label{sect:3}

Imagine the following experiment. We have a 2D photonic system with an implemented synthetic magnetic field 
and the Kerr type nonlinearity. We launch a beam with some initial profile into this system and 
ask whether the trajectory from this initial condition is stable or not. In the paraxial approximation 
this system is modeled with the NLSE. 
An equivalent experiment in BECs would be to prepare the weakly interacting BEC in some initial state, 
in a potential confining the dynamics to two dimensions;
then, we suddenly turn on the synthetic magnetic field, and wonder whether 
subsequent dynamics is sensitive to small perturbations on the initial state. 

This is modeled with the NLSE in 2D with an additional vector potential $\bf{A}$,
\begin{equation}
i \frac{\partial \psi}{\partial t} = \left[ \left(-i \nabla - \bf{A} \right)^2 + \eta \left| \psi \right|^2 \right] \psi, 
\label{nlse_2d}
\end{equation}
where $\nabla=\hat x \frac{\partial}{\partial x}+\hat y \frac{\partial}{\partial y}$ and $\psi \equiv \psi(x,y,t)$. 
In photonics, the ''time variable'' is the propagation axis coordinate $z$ instead of $t$~\cite{Ozawa2019}. 
The vector potential $\bf{A}$ corresponds to a homogeneous synthetic magnetic field perpendicular to the 
2D plane, ${\bf B} = B \hat z = \nabla \times {\bf A}$. 
We should have the freedom to choose a gauge for the vector potential $\bf{A}$
and focus on two most common choices, the symmetric and the Landau gauge.

In the symmetric gauge ${\bf A}_S = \frac{1}{2} B (- y \hat x + x \hat y)$, which leads to the following time-dependent NLSE in 2D,
\begin{align}
i \frac{\partial \psi}{\partial t} =& \left[ - \frac{\partial^2}{\partial x^2} - \frac{\partial^2}{\partial y^2} 
								-i B y \frac{\partial}{\partial x} + iBx \frac{\partial}{\partial y} \right. \nonumber \\
							&+ \left. \frac{1}{4} B^2 \left( x^2 + y^2 \right)
								+\eta \left| \psi \right|^2 \right] \psi.
\label{symm}
\end{align}
We can write the Landau gauge either as ${\bf A}_{Lx} = -B y \hat x$ or ${\bf A}_{Ly} = B x \hat y$. 
With the choice ${\bf A}_{Lx}$, the time-dependent NLSE reads
\begin{align}
i \frac{\partial \psi}{\partial t} =& \left( - \frac{\partial^2}{\partial x^2} - \frac{\partial^2}{\partial y^2} 
								-2 iBy \frac{\partial}{\partial x} + B^2  y^2+\eta \left| \psi \right|^2 \right) \psi,
\label{Landaux}
\end{align}
and with ${\bf A}_{Ly}$ we have
\begin{align}
i \frac{\partial \psi}{\partial t} =& \left( - \frac{\partial^2}{\partial x^2} - \frac{\partial^2}{\partial y^2} 
								+2 iBx \frac{\partial}{\partial y} + B^2  x^2+\eta \left| \psi \right|^2 \right) \psi.
\label{Landauy}
\end{align}
It is obvious that dynamics from an identical initial condition $\psi(x,y,t=0)$ 
will differ if we evolve it in a different gauge, i.e., evolution with 
Eqs. (\ref{symm}), (\ref{Landaux}), or (\ref{Landauy}) will differ. 
However, this does not mean that the gauge invariance principle is violated.
Equations (4), (5), and (6) are related through gauge transformations of the vector potential. However, in order for all three of them to yield the same dynamics, the initial condition for one of them should be $\psi(x,y,0)$, and the initial conditions for the other two should be appropriately gauge transformed. Because the system is experimentally prepared in the state $\psi(x,y,0)$ at $t=0$, it is not immediately clear which equation, that is, which gauge to use for this initial condition. 

To understand what happened, note that in the gedanken experiments described above, in the photonic and BEC 
context, the system is prepared in some state, and then at $t=0$ the synthetic magnetic field is suddenly turned on 
(in photonics this is the moment the optical beam enters the 2D medium). 
Thus, $B(t)=B\theta(t)$, and the vector potential ${\bf A}(x,y,t)={\bf A}(x,y)\theta(t)$, where $\theta(t)$ is Heaviside step function. 
By Faraday's law, the synthetic magnetic field at the instant $t=0$ creates a 
spatially dependent synthetic electric field kick, 
which differs for the gauges mentioned above.
We obtain the following electric fields in different gauges:
${\bf E}_S = -\frac{\partial {\bf A}_S}{\partial t}= - B \delta(t) \left( -\frac{y}{2} \hat x +\frac{x}{2} \hat y \right)$,
${\bf E}_{Lx} = -\frac{\partial {\bf A}_{Lx}}{\partial t}=  B \delta(t) y \hat x$, and
${\bf E}_{Ly} = -\frac{\partial {\bf A}_{Ly}}{\partial t}=  - B \delta(t) x \hat y$.
Thus, even though at times $t>0$ the fields generated by the vector potential are identical,
this kick affects dynamics for times $t>0$. 

One can ask next, which gauge should we use for a given experiment? 
This depends on the experiment and the way synthetic magnetic field is implemented at $t=0$. 
A given implementation of the uniform synthetic magnetic field will have a particular and unique 
synthetic electric field kick at $t=0$. The gauge used for the dynamics should yield 
exactly the same synthetic electric field kick in order to describe the experimental situation. In the next section we will discuss the MI phenomenon in the aforementioned different gauges, which 
as we have just explained correspond to different experimental implementations of the field. 
Before that let us elaborate how we choose the initial condition and explore MI.

When we introduce the vector potential into NLSE, new terms on the r.h.s of \cref{symm,Landaux,Landauy} appear: first-order spatial partial derivatives and harmonic confinement terms. 
Due to the harmonic terms, we do not proceed with the standard MI analysis by using plane wave 
as an initial state, because plane waves are not eigenstates of neither of \cref{symm,Landaux,Landauy}.
However, following the discussion from the previous section, for the initial state we choose 
the ground state of the 2D NLSE with isotropic harmonic confinement $ \frac{1}{4} B^2 \left( x^2 + y^2 \right)$
\begin{align}
	\mu \psi =& \left[ - \frac{\partial^2}{\partial x^2} - \frac{\partial^2}{\partial y^2} 
					+ \frac{1}{4} B^2 \left( x^2 + y^2 \right)
					+\eta' \left| \psi \right|^2 \right] \psi,
\end{align}
in the TF approximation. We choose $\mu=1$ for the chemical potential, the field is set to $B=0.1$, and the strength of 
the nonlinearity is positive, $\eta'=1$. In the TF approximation we neglect the kinetic energy terms, which leads to 
the ground state wave function
\begin{equation}
\psi_{TF}(x,y) = \sqrt{\frac{\mu-\frac{1}{4}B^2 \left( x^2 + y^2 \right)}{\eta'}}
\label{tf_state}
\end{equation}
for $0<\sqrt{x^2+y^2}<2 \frac{\sqrt{\mu}}{B}$, and $\psi_{TF}=0$ elsewhere. 
We observe the time-evolution of the system from this initial state; we propagate it with  
Eq. \eqref{nlse_2d} with nonlinearity $\eta = -\eta'$, because the MI is expected in the regime of negative nonlinearities
(attractive BECs or self-focusing nonlinear media).

\section{Modulation instability in 2D NLSE with a synthetic magnetic field}
\label{sect:4}

\begin{figure}
\includegraphics[width=0.48\textwidth,bb=330 0 795 511, clip=true]{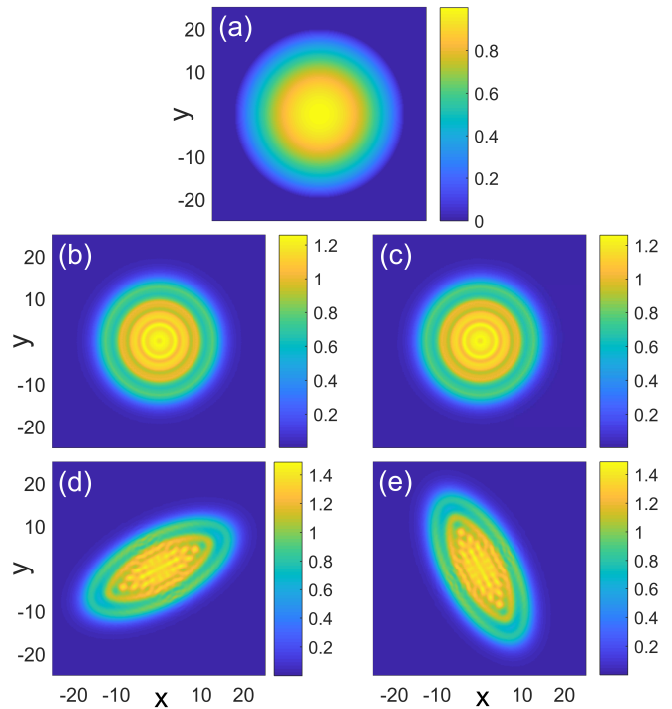}
\caption{(a) Density $|\psi_{00}(x,y,0)|^2=|\psi_{TF}|^2$ of the initial Thomas-Fermi state \eqref{tf_state}. Density $|\psi_{00}(x,y,t=3)|^2$ evolved with 
(b) the harmonic potential only [Eq. \eqref{nlse2d_harm}], 
(c) the symmetric gauge ${\bf A}_S$ [Eq. \eqref{symm}], 
(d) the Landau gauge ${\bf A}_{Lx}$ [Eq. \eqref{Landaux}], and 
(e) the Landau gauge ${\bf A}_{Ly}$ [Eq. \eqref{Landauy}].}
\label{fig:Fig2}
\end{figure}

In this section, we present numerical results which demonstrate MI in our system.
For this, we have implemented a 2D split-step method for the time-evolution with the NLSE \eqref{nlse_2d} 
in the symmetric and the Landau gauge, which include the first-order spatial partial derivatives 
and harmonic terms arising from the vector potential.
For comparison, we will also show results for the time-evolution of the NLSE with only the harmonic terms present, that is, 
for 2D generalization of Eq. \eqref{nlse1d},
\begin{align}
i \frac{\partial \psi}{\partial t} =& \left[ - \frac{\partial^2}{\partial x^2} - \frac{\partial^2}{\partial y^2} 
								+ \frac{1}{4} B^2 \left( x^2 + y^2 \right)
								+\eta \left| \psi \right|^2 \right] \psi.
\label{nlse2d_harm}
\end{align}
The initial state $\psi_{00}(x,y,0)=\psi_{TF}$ given in \eqref{tf_state} is illustrated in Fig. \ref{fig:Fig2}(a). 
The time-evolution of this state propagated in different gauges [Eq. \eqref{nlse2d_harm} and \cref{symm,Landaux,Landauy}] 
is presented in Figs. \ref{fig:Fig2}(b)-(e).
%
%
We observe that small density modulations develop in time because the initial TF state is not an eigenstate 
of the system in any gauge.
The time-evolution with only the harmonic term present \eqref{nlse2d_harm}, and with the magnetic field in the 
symmetric gauge \eqref{symm} are hardly distinguishable for the chosen set of parameters [compare Figs. \ref{fig:Fig2}(b) and (c)].
In the Landau gauge(s) the density cloud becomes elongated and rotated in time [Figs. \ref{fig:Fig2}(d,e)]. 
The two density clouds in Figs. \ref{fig:Fig2}(d) and (e) are oriented at $\pi/2$ angle relative to each other, 
which reflects the geometric relationship between Landau gauges ${\bf A}_{Lx}$ and ${\bf A}_{Ly}$.

We now add perturbations to the initial state,
\begin{equation}
\psi_{k_x k_y}(x,y,0)=\mathcal{N}_{k_xk_y}\psi_{TF}\left[1+0.01 \cos \left(k_x x+k_y y\right)\right].
\label{tf_state_kxky}
\end{equation}
Here, ${\bf k}=(k_x,k_y)$ is a 2D momentum of the perturbation, and $\mathcal{N}_{k_x, k_y}$ is the normalization constant such that perturbed TF states have the same normalization as the unperturbed TF state \eqref{tf_state}.
In order to consistently compare time-evolution of the perturbed \eqref{tf_state_kxky} and unperturbed initial state \eqref{tf_state} we show 
time-evolved density profiles at $t=3$.

\begin{figure}
\includegraphics[width=0.48\textwidth,bb=330 0 795 511, clip=true]{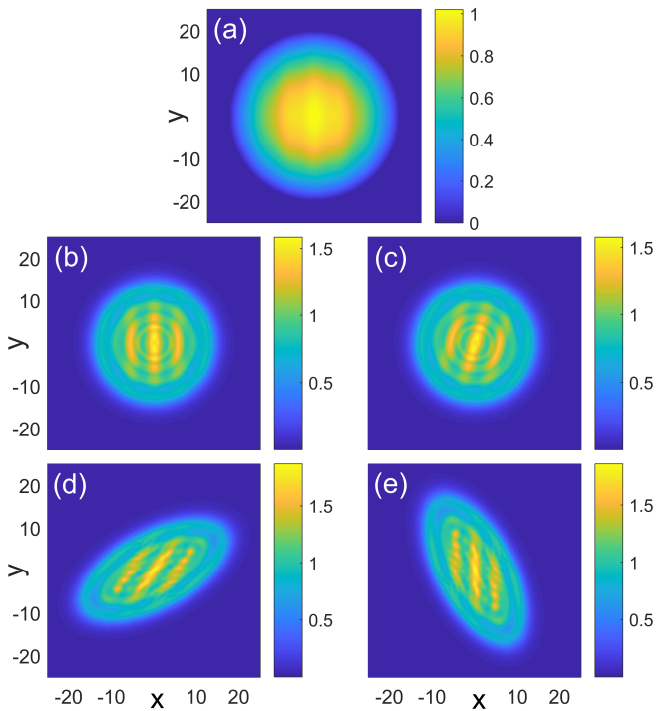}
\caption{(a) Density $|\psi_{10}(x,y,0)|^2$ of initial state \eqref{tf_state_kxky} with perturbation momentum $(k_x=1,k_y=0)$. (b)-(e) Density $|\psi_{10}(x,y,t=3)|^2$ evolved with the same equations as in the Fig. \ref{fig:Fig2}(b)-(e).}
\label{fig:Fig3}
\end{figure}

First, we consider $\psi_{10}(x,y,0)$ with perturbation momentum $(k_x,k_y)=(1,0)$.
The density profile of this initial state is shown in Fig. \ref{fig:Fig3}(a), and its time-evolution 
in different gauges in Figs. \ref{fig:Fig3}(b)-(e); the outline of Figs. \ref{fig:Fig2} and \ref{fig:Fig3} are identical for easier comparison. 
%
%
We see that in all gauges the MI is present, as can be seen from the strong modulations in the densities.
Time-evolution with the harmonic confinement only [Eq. \eqref{nlse2d_harm}], and with the 
symmetric gauge [Eq. \eqref{symm}] show MI with the same intensity modulations [Figs. \ref{fig:Fig3}(b) and (c)],
however, in the symmetric gauge we observe rotation of modulation patterns.
In the Landau gauge ${\bf A}_{Lx}$ (Eq. \eqref{Landaux}), both the density cloud and its modulation patterns are elongated and rotated in time (Fig. \ref{fig:Fig3}(d)).
In contrast, in the Landau gauge ${\bf A}_{Ly}$ [Eq. \eqref{Landauy}], the cloud is also elongated and rotated, but modulation patterns do not rotate during the time-evolution [Fig. \ref{fig:Fig3}(e)]. This difference in the modulation patterns in Landau gauges reflects the differences between synthetic electric field kicks at $t=0$ for ${\bf A}_{Lx}$ and ${\bf A}_{Ly}$, i.e., synthetic electric field ${\bf E}_{Lx}$ provides initial momentum perpendicular to the stripes of perturbation in $\psi_{10}(x,y,0)$, while synthetic electric field ${\bf E}_{Ly}$ provides initial momentum parallel to the stripes of that perturbation, which results in different dynamics of the modulation pattern later on.      

\begin{figure}
\includegraphics[width=0.48\textwidth,bb=330 0 795 511, clip=true]{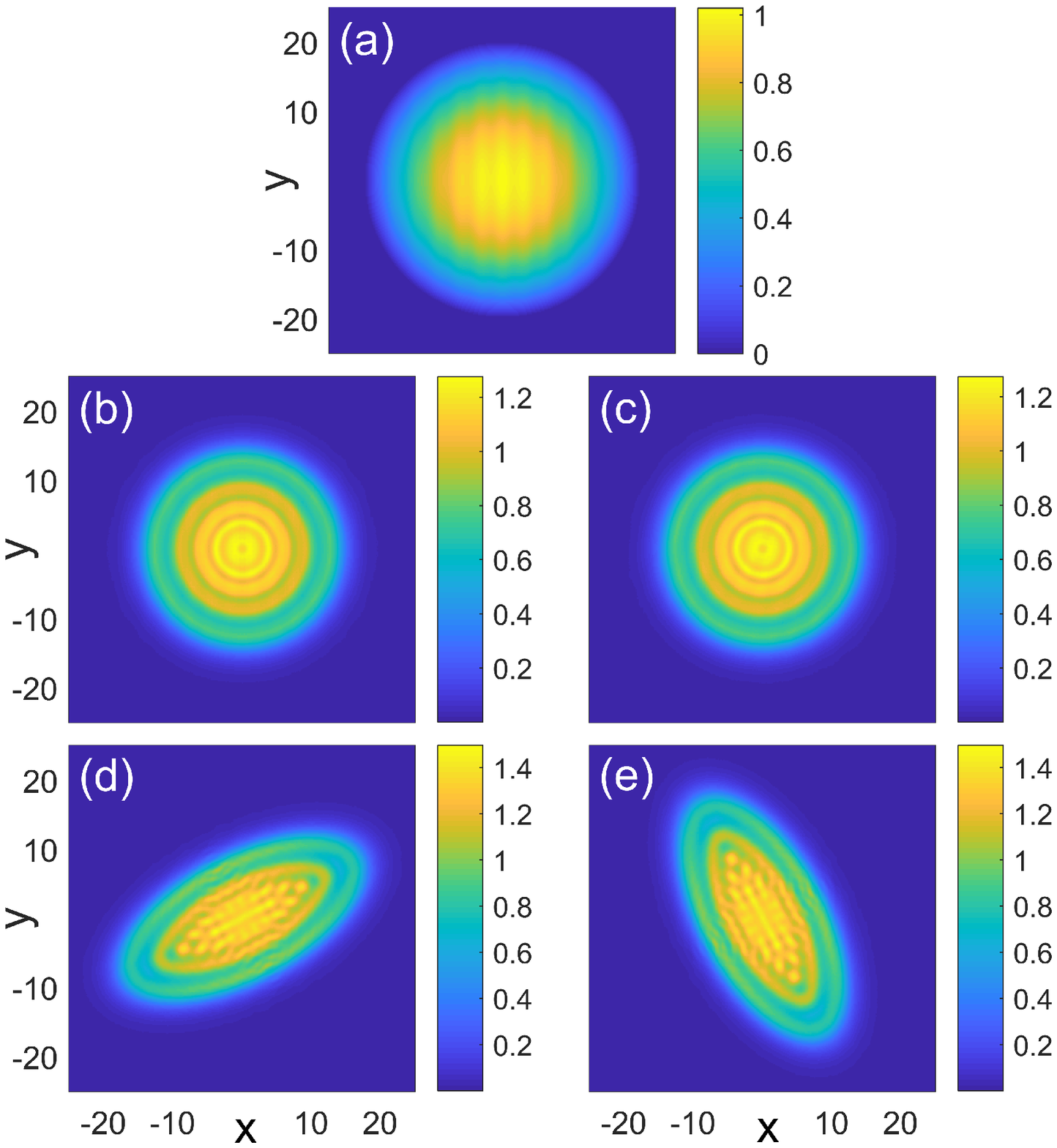}
\caption{(a) Density $|\psi_{20}(x,y,0)|^2$ of initial state \eqref{tf_state_kxky} with perturbation momentum $(k_x=2,k_y=0)$. (b)-(e) Density $|\psi_{20}(x,y,t=3)|^2$ evolved with the same equations as in the Fig. \ref{fig:Fig2}(b)-(e).}
\label{fig:Fig4}
\end{figure}

Second, we consider time-evolution from the initial state $\psi_{20}(x,y,0)$ with perturbation momentum $(k_x,k_y)=(2,0)$.
Results are presented in Fig. \ref{fig:Fig4}; the outline of the figures is again identical to those in Figs. \ref{fig:Fig2} and \ref{fig:Fig3}.
%
%
This initial perturbation does not destabilize the trajectory, i.e., we do not observe MI in any gauge.

\begin{figure}
\includegraphics[width=0.48\textwidth,bb=330 0 795 511, clip=true]{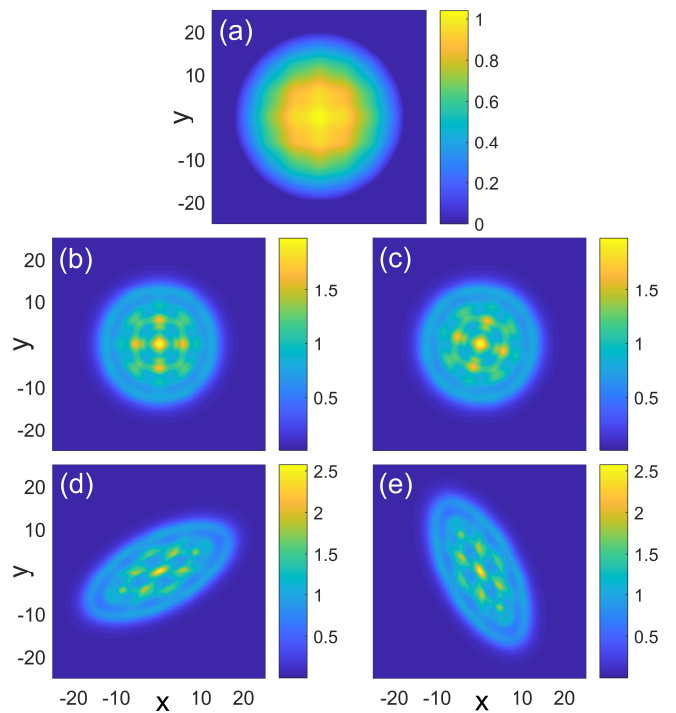}
\caption{(a) Density $|\tilde{\psi}_{11}(x,y,0)|^2$ of the perturbed initial state \eqref{tf_state3}. (b)-(e) Density $|\tilde{\psi}_{11}(x,y,t=3)|^2$ evolved with the same equations as in the Fig. \ref{fig:Fig2}(b)-(e).}
\label{fig:Fig5}
\end{figure}

In a more general setting, perturbation may consist of two or more components in momentum space.
As an example, we investigate evolution of the TF initial state,
\begin{equation}
\tilde{\psi}_{11}(x,y,0) =\tilde{\mathcal{N}}_{11} \psi_{TF}  \left\{1 + 0.01 \left[\cos(x) + \cos(y) \right] \right\},
\label{tf_state3}
\end{equation}
where $\tilde{\mathcal{N}}_{11}$ is the normalization constant determined as in Eq. \eqref{tf_state_kxky}.
The initial state \eqref{tf_state3} is a superposition of perturbations with momenta 
$(k_x,k_y)=(1,0)$ and $(k_x,k_y)=(0,1)$. 
The MI is revealed during time evolution for this initial state [see Fig. \ref{fig:Fig5}], as expected from the results for 
a single unstable $(k_x,k_y)=(1,0)$ perturbation. 
The modulation patterns which develop during dynamics have more complex shapes than the simple modulation 
stripes from Fig. \ref{fig:Fig3}. 
Here, density modulations form a lattice when only the harmonic term is present (Fig. \ref{fig:Fig5}(b)), 
and this lattice is rotated in the symmetric gauge (Fig. \ref{fig:Fig5}(c)).
%
%
In addition to this, in both Landau gauges, the atomic cloud is elongated and rotated, which leads to nontrivial MI density patterns (Fig. \ref{fig:Fig5}(d,e)).

\begin{figure}
\includegraphics[width=0.48\textwidth]{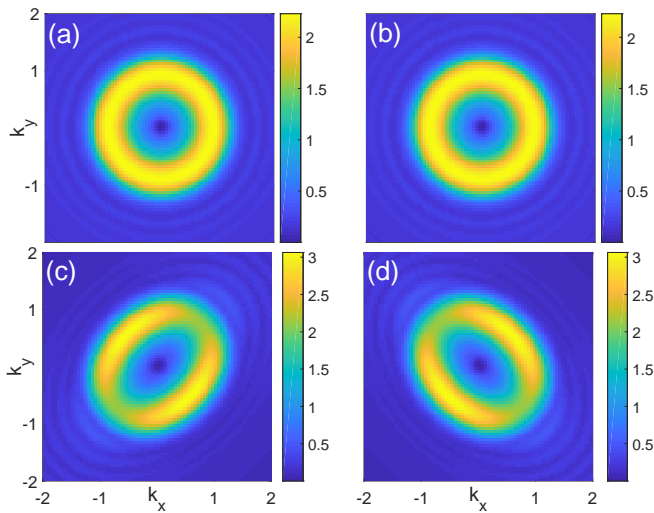}
\caption{Instability region in the momentum space for the time evolution with: 
(a) the harmonic potential only [Eq. \eqref{nlse2d_harm}], 
(b) the symmetric gauge ${\bf A}_S$ [Eq. \eqref{symm}], 
(c) the Landau gauge ${\bf A}_{Lx}$ [Eq. \eqref{Landaux}], and 
(d) the Landau gauge ${\bf A}_{Ly}$ [Eq. \eqref{Landauy}]. See text for details.}
\label{fig:Fig6}
\end{figure}

In order to have a dynamical measure of the MI phenomenon which emerges during time evolution, 
and numerically investigate the instability region in the momentum space of perturbations, we introduce the following quantity:
\begin{equation}
\Gamma_{k_x,k_y}(t)=\sqrt{\int \left[ \left|\psi_{k_xk_y}(x,y,t)\right|^2-\left|\psi_{00}(x,y,t)\right|^2\right]^2 dx dy}.
\label{Delta}
\end{equation}
In Fig. \ref{fig:Fig6} we have calculated $\Gamma_{k_x,k_y}(t=3)$ after time-evolution with Eq. \eqref{nlse2d_harm} and \cref{symm,Landaux,Landauy}.    
%
%
The (in)stability region in momentum space for evolution with the harmonic potential only [Eq. \eqref{nlse2d_harm}] 
and in the symmetric gauge ${\bf A}_S$ [Eq. \eqref{symm}] are presented in Figs. \ref{fig:Fig6}(a) and (b), respectively. 
Results are indistinguishable; in both cases $\Gamma_{k_x,k_y}$ has azimuthal symmetry, with 
the maximally unstable perturbation at radius $k =\sqrt{k_x^2+k_y^2} \approx 0.95$.
In Figs. \ref{fig:Fig6}(c) and (d), we show the (in)stability region for time-evolution with the Landau gauges 
${\bf A}_{Lx}$ [Eq. \eqref{Landaux}] and ${\bf A}_{Ly}$ [Eq. \eqref{Landauy}], respectively.
These instability regions are of elliptical shape and perpendicular to each other.
The maximally unstable perturbations in Fig. \ref{fig:Fig6}(c) are at ${\bf k}\approx (-0.55,0.55)$ and $(0.55,-0.55)$;
we can also see a drop of instability in the vicinity of ${\bf k}\approx (0.75,0.75)$ and $(-0.75,-0.75)$.

More complex, direction dependent behavior of the instability region for the Landau gauges is 
attributed to the difference in the symmetries between the two Landau gauges and the initial TF state.
The switching of the synthetic magnetic field in the Landau gauge introduces an electric field with translational symmetry;
thus, it provides the momentum kick which breaks the cylindrical symmetry of the initial TF state.
This further leads to the distortion of the density cloud in time, which is reflected in the momentum instability region as well.
When the synthetic magnetic field is turned on in the symmetric gauge, it generates an electric field with 
the cylindrical symmetry; thus, the symmetry of the initial TF state is preserved during the time-evolution, 
both in the real and the momentum space.

\section{Conclusion}
\label{sect:5}

In conclusion, we have explored the phenomenon of modulation instability for 2D systems which obey the NLSE in 
a homogeneous synthetic magnetic field.
We have explored MI for the trajectory evolving from the initial state which has a Thomas-Fermi profile \eqref{tf_state}. 
Small perturbations upon the initial state were characterized in momentum space [Eq. \eqref{tf_state_kxky}]. 
Some perturbations destabilize the trajectory, others do not, as expected. 
We have calculated the region of (in)stability in momentum space, as presented in Fig. \ref{fig:Fig6}. 
The stability depends on the gauge used, however, this does not mean that gauge invariance is violated. 
We have pointed out that when the synthetic magnetic field is turned on, 
there will be an instantaneous electric field kick to the system arising from Faraday's law, 
which depends on the gauge used; all gauges yield identical fields for $t>0$. 
When an experiment is theoretically simulated, the gauge should be chosen to properly describe this 
initial synthetic electric field kick. 
We envision that our results will prove useful in studying instabilities that appear either in experiments with ultracold 
atomic gases, or when studying light propagation in nonlinear media, both with synthetic magnetic fields.

\section{Acknowledgments}

This work was supported by the Croatian Science Foundation Grant No. IP-2016-06-5885 SynthMagIA, and in part by the QuantiXLie Center of Excellence, a project co-financed by the Croatian Government and European Union through the European Regional Development Fund - the Competitiveness and Cohesion Operational Programme (Grant KK.01.1.1.01.0004).


\end{document}